# Skills-based on technological knowledge in the digital economy activity

Dr. Cesar R Salas-Guerra

*EDUCTUM Business School, Miami Lakes, United States of America*

*UAB, Universitat Autónoma de Barcelona, Spain*

*cesar.salas@eductum.us*

*cesar.salasg@e-campus.uab.cat*

*ORCID ID: 0000-0001-7182-3002*

# Abstract

This research seeks to measure the impact of people with technological knowledge on regional digital economic activity and the implications of prosperous cities' contagion effect on neighbouring ones. The focus of this study is quantitative, cross-sectional, and its design is correlational-causal. This study covers seven micro-regions of Minas Gerais in Brazil, organized in 89 municipalities, with 69% urban population and 31% rural. The data used consisted of 4,361 observations obtained in the Brazilian government's public repositories, organized into panel data, and analysed using partial least squares, micro-regional spatial regression, and identification patterns with machine learning. The confirmatory analysis of the regression test establishes a significant impact between the CE's technological knowledge and the digital economic activity AED through a predictive value of $R2 = .749$, $\beta = .867$, $p = .000$ (value t = 18,298). With high notoriety among the variables, public and private university institutions (IUPP), professors with doctorates and masters (DCNT), and information technology occupations (CBO). A geographic concentration of companies that demand technology-based skills had effects by slowing down the development of small municipalities, suggesting the development of new government technology initiatives that support new business models based on technological knowledge.

## Introduction

Economic theories say that the level of production increases due to the transfer of knowledge in the human capital and information technologies (Cvetanović et al., 2015); this has led to the development of new production processes based on digitization and data management (Agrawal et al., 2019).

With COVID-19, companies face many challenges; however, for years, the new characteristics of human digital adaptation (Engen et al., 2017) have promoted new business models focused on new technologies (Boisier, 2016; Espino-Timón, 2017), favouring diversification of regional economies (Boschma, 2017).

These new business models use technologies based on information architectures (Paea & Baird, 2018) with the use of artificial intelligence (AI) algorithms and predictive analysis models (Agrawal et al., 2019), minimizing in this way dependence on technological externalities (Ma & Zhao, 2019).

Consequently, the knowledge generated by technological innovation and the transformation of traditional economic activity to digital (Mcadam & Mcadam, 2016; Vega-Gomez et al., 2018) has driven the need for skilled labour in technological innovation (Grillitsch & Trippl, 2018). However, some regions may be affected by the sudden agglomeration of prosperous cities (Nagy et al., 2018a), which could polarize economic growth and regional development (Ma & Zhao, 2019).

Therefore, based on the problem previously exposed, we propose the following research question: What impact does technological knowledge have on digital economic activity, and what are the effects of business agglomeration? Next, we will describe the study variables.

**Classification and definition of the study variables**

This study's variables are made up of constructs and indicators, described below according to the order they were considered in the conceptual model.

Table 1. Variables definition matrix

| Code | Variable | Definition | References |
|---|---|---|---|
| **CE** | **Technological knowledge** | It comprises higher-level institutions with great importance in regional economic development and knowledge individuals related to technological, scientific knowledge to create and capture value for a company. | (Giones & Brem, 2018; Bialetti, 2012; Ogundari & Awokuse, 2018; Cortright, 2001) |
| **AED** | **Digital economic activity** | It can generate productivity, economic growth, and well-being through regional economies' construction based on innovation and technological knowledge. | (Brynjolfsson et al., 2018; Mendez, 2016; Cortright, 2001; Gomez, 2017; Manuel et al., 2016; dos Santos, 2017; Century et al., 2007) |

*Construct technological knowledge (CE)*

- Number of public and private university institutions (IUPP) that comprise institutions that offer professional technical education, technological education institutions, educational institutions of undergraduate education, educational institutions of graduate education, and educational institutions of comprehensive education by the municipality between 2009 and 2018.

- Number of teachers (DCNT) with doctorates and masters, professional studies instructors, computer science professors, science professors, economics and administrative science professors, biological sciences professors, engineering professors, mathematics and statistics professors, human sciences professors of public and private university institutions by the municipality between 2009 and 2018;
- Amount of government economic contribution in the municipal education system (AEF) between 2009 and 2018.

*Construct digital economic activity (AED)*

- Number of IT companies per municipality (ICT) between 2009 and 2018.
- Brazilian classification of occupations (CBO) composed by number of network analysts data communication, information technology services manager, information technology production manager, information technology project manager, information technology technical support manager, database administrator, electrical and telephone line maintenance electrician data communication, data communication and telephone network installer-repairer, data communication technician, information security administrator, IT services manager, information analyst (information network researcher), IT project manager Information Technology Production Manager, Information Technology Technical Support Manager training, information systems programmer, computer equipment maintenance technician, computer user support technician, information technology management technologist, computer programming engineer, computer hardware engineer, computer operating systems engineer , installer of electronic equipment (computers and auxiliary equipment), computer operator (including microcomputer), internet programmer, all by municipality between 2009 and 2018.

**Theoretical framework**

*Endogenous growth theory*

This economic theory incorporates two essential points; First, it describes technological progress as a product of economic activity; second, this theory holds that unlike physical objects, knowledge and technology contribute to economic growth (Cortright, 2010).

Although the neoclassical school affirms that the growth of the value of production at the regional level is the result of an increase in the quantity and quality of labour, for the new growth theory, the nature of the change is based on the technological factor (Cvetanović et al., 2015).

The technological level of production increases through the transfer of knowledge and innovation activities through the contribution of innovative educational systems (Giones & Brem, 2018). Endogenous economic growth develops at a rate determined by the economic system, mainly those that provide stimuli for the creation of technological knowledge (Howitt, 2010). This theoretical basis is reflected by the following equation:

$$y = AK \qquad (1)$$

Where, A should be understood as an expression that represents technological factors, while K includes human capital as knowledge (Pack, 1994). The model most used in the theoretical and empirical analysis of growth and productivity over many years was the Cobb-Douglas function, according to Adams (2005); because the estimation of the parameters as seen in Equation (1) is fundamental in input determinants specific such as knowledge, labour skills, the distinction between capital

(ICT), non-capital (ICT) and the factor of technological progress (Hanclova et al., 2015).

Akcigit (2013) found that regions far from the technological frontier enjoy a "lagged advantage"; This disadvantage implies that, in the long term, a region with a low rate of innovation will lag (Paunov, 2017) due to competitiveness in regional technological development.

**The new industrial revolution**

The industry is known as the economic process for highly automated material goods (Lasi et al., 2014). However, the fourth industrial revolution has transformed production thanks to technology in production processes; it includes, but is not limited to, artificial intelligence and the internet of things (Vasin et al., 2018).

The flow of data managed locally and remotely between the different elements of this new value creation ecosystem called "Industry 4.0" (Stock & Seliger, 2016) allows companies to connect their machinery in the physical-cybernetic spectrum, for this unleash actions through the exchange of information autonomously (Du et al., 2018).

Industry 4.0, as the fourth industrial revolution is known, has allowed the developing an environment where all the elements coexist together continuously and ubiquitously (Zambon et al., 2019). Although computers and automation have existed since previous decades, this interaction process seeks to establish coordination capacities to improve product management through the correct use of data (Nagy et al., 2018b).

Consequently, recent studies found that the pressure of competition towards organizations motivates companies to trust more in knowledge and data use (Manesh et al., 2019), seeking to maintain a long-term competitive advantage, thus generating a priority for many research university's (Du et al., 2018) that support professionals in the development of appropriate solutions.

**Technological knowledge**

The current literature review establishes that technological knowledge brings together people closely related to scientific knowledge and innovation, which contribute to the development of digital economic activity (Giones & Brem, 2018), notably establishing academic institutions' role in regional economic development (Feliu & Rodríguez, 2017).

      As a basis for regional development, Vega-Gomez et al. (2018) mention that technological knowledge seeks at the regional level to alter economic structures through a) innovation, b) detection of opportunities and c) business creation; thus producing knowledge providers, sophisticated demand, skilled labour, educational research activities, and support services, thus creating a competitive advantage (Grillitsch & Trippl, 2018).

      Consequently, the impetus that academic institutions exercise in economic development through university technology (Mcadam & Mcadam, 2016) implies a high range of university-community knowledge exchange. This new business approach based on new technologies is also known as academic spin-offs (Vega-Gomez et al., 2018), which positively affect innovation processes and business technology entrepreneurship (Li & Nuccialleri, 2016).

      Finally, the review of the literature recognizes that there is a direct relationship between business creation and economic growth, as mentioned by Lupiáñez Carrillo et al. (2017) where they affirm that the fundamental factor of economic growth is innovation through digital economic activity, which is driven by technological knowledge.

The discussion of the literature presented above allows us to propose the hypothesis of this research:

H1: Technological knowledge drives digital economic activity.

**Digital Economic Activity**

The digital revolution, also known as the digital economy (Li & Nuccialleri, 2016), has affected how conventional markets for products and services compete through the transition to the information economy. According to Gutiérrez et al. (2016), since through innovation, companies optimize a profit function and establish models of search and selection of technologies.

The literature review establishes that digital economic activity can generate productivity, economic growth, and well-being through regional economies' construction based on innovation and knowledge (Brynjolfsson et al., 2018; Mendez, 2016). Similarly, it establishes that economies' competitiveness depends on the use of new technologies, the same ones that contribute to the gross domestic product (Neffati & Gouidar, 2019).

Information technology companies have been characterized lately by their influence on production (Giones & Brem, 2018) marketing processes, job creation (Neffati & Gouidar, 2019), the transformation of the workforce, and business innovation through the development of new business models based on digital platforms (Agrawal et al., 2019), which maximize business management (Lu et al., 2018).

In conclusion, the promotion of technology transfer through technological knowledge by information technology companies solves the need to build regional economies based on innovation and knowledge (Mendez, 2016), since innovation through the economy of digital activity (Lupiáñez et al., 2017) is one of the fundamental causes of economic growth (Kumar & Dahiya, 2016).

**Methodology**

*Methodological design of the research*

The focus of this research is quantitative, cross-sectional since the variables of interest of a specific population were studied over time using multivariate spatial analysis statistics (Gomez, 2017) and structural equations with partial least squares (Ajamieh, 2016) through the implementation of a matrix of spatial weights to determine the impact relationships of the digital economic activity of a municipality in comparison with its neighbor.

The methodology was framed in the cross-sectional correlational-causal design because only the level of correlation between the variables was measured to identify possible causalities in the phenomenon that will later be studied (Orengo, 2008); The data used consisted of 4,361 observations obtained in the Brazilian government's public repositories described below:

- Brazilian Classification of Occupations (CBO)
- National Classification of Economic Activities (CNAE)
- Brazilian Institute of Geography and Statistics (IBGE), and
- Central Bank of Brazil (BCB)

The panel data allowed identifying systematic and unobserved differences between the units correlated with factors whose effects should be measured (Wooldridge, 2009). They made it possible to generalize the results since this study seeks to obtain from this population the data previously organized in tables methodologically designed for such purposes (IBGE, 2017).

**Geographical projection**

Brazil's geographical projection is composed of a hierarchical distribution (Bosco, 2009), where the whole country represents the first level of the hierarchy, followed by the second level composed of the north region, northeast region, middle west southeast, and southern region. The third level has twenty-seven states or administrative units. One hundred and thirty-seven mesoregions represent the fourth level, and the fifth level corresponds to the microregions, which are subdivisions of the mesoregions; the country has 558 microregions (IBGE, 2019).

Finally, the sixth level corresponds to the municipalities, which are made up of 5,565. However, in reality, two have a special status: the Federal District, which is considered a unit at all hierarchical levels, that is, it is a State, a mesoregion, a microregion, and a municipality at the same time, and the archipelago of Fernando de Noroña, which is a state district, belonging to Pernambuco, but considered as a municipality for analysis purposes.

This study covers the total population composed of 7 microregions organized by 89 municipalities that comprise 128,602 km², equivalent to 22% of the state of Minas Gerais, with a population of 1,610,413 inhabitants. 69% is urban, and 31% rural (IBGE, 2017).

Table 2. Municipalities of the northern mesoregion of Minas Gerais

| Microregion | Municipalities | Municipalities | Municipalities |
|---|---|---|---|
| 1. Bocaiuva | **Bocaiúva *** | Engenheiro Navarro<br>Francisco Dumont | Guaraciama<br>Olhos de Água |

| Region | | | |
|---|---|---|---|
| 2. Grão-Mogol | Botumirim | **Grão-Mogol *** | Josenópolis |
| | Cristália | Itacambira | Padre Carvalho |
| 3. Janaúba | Catuti | Mamonas | Porteirinha |
| | Espinosa | Mato Verde | Riacho dos Machados |
| | Gameleiras | Monte Azul | Serranópolis de Minas |
| | Jaíba | Nova Porteirinha | |
| | **Janaúba *** | Pai Pedro | |
| 4. Januária | Bonito de Minas | São João das Missões | Montalvânia |
| | Chapada Gaúcha | Urucuia | Pedras de Maria da Cruz |
| | Cônego Marinho | Juvenília | Pintópolis |
| | Icaraí de Minas | Manga | São Francisco |
| | Itacarambi | Matias Cardoso | |
| | **Januária *** | Miravânia | |
| 5. Montes Claros | Brasília de Minas | Ibiracatu | Patis |
| | Campo Azul | Japonvar | Ponto Chique |
| | Capitão Enéias | Juramento | São João da Lagoa |
| | Claro dos Poções | Lontra | São João da Ponte |
| | Coração de Jesus | Luislândia | São João do Pacuí |
| | Francisco Sá | Mirabela | Ubaí |
| | Glaucilândia | **Montes Claros *** | Varzelândia |
| | | | Verdelândia |
| 6. Salinas | Águas Vermelhas | Ninheira | Santo Antônio do Retiro |
| | Berizal | Novorizonte | São João do Paraíso |
| | Curral de Dentro | Rio Pardo de Minas | Taiobeiras |
| | Divisa Alegre | Rubelita | Vargem Grande do Rio Pardo |
| | Fruta de Leite | **Salinas *** | Montezuma |
| | Indaiabira | Santa Cruz de Salinas | |
| 7. Pirapora | Buritizeiro | Lassance | Santa Fé de Minas |
| | Ibiaí | **Pirapora *** | São Romão |
| | Jequitaí | Riachinho | Várzea da Palma |
| | Lagoa dos Patos | | |

Note: * Region capital

**Data Analysis**

In this study, 4,361 observations were analysed, organized into panel data; the process and tools are detailed below:

- The first analysis phase: reflective PLS model (Smart PLS 3.0)

- The second analysis phase: microregional spatial regression (Geoda 1.6 and ArcGIS Pro)
- The third analysis phase: municipal growth analysis patterns by variable (BigML's platform for Machine Learning)

*First analysis phase: reflective PLS model*

For this first phase, a non-parametric reflective model of partial least squares PLS and Bootstrapping is used since it is reliable and less sensitive to outliers. The model consists of two constructs, and fifty indicators previously explained.

Figure 1. Research empirical model

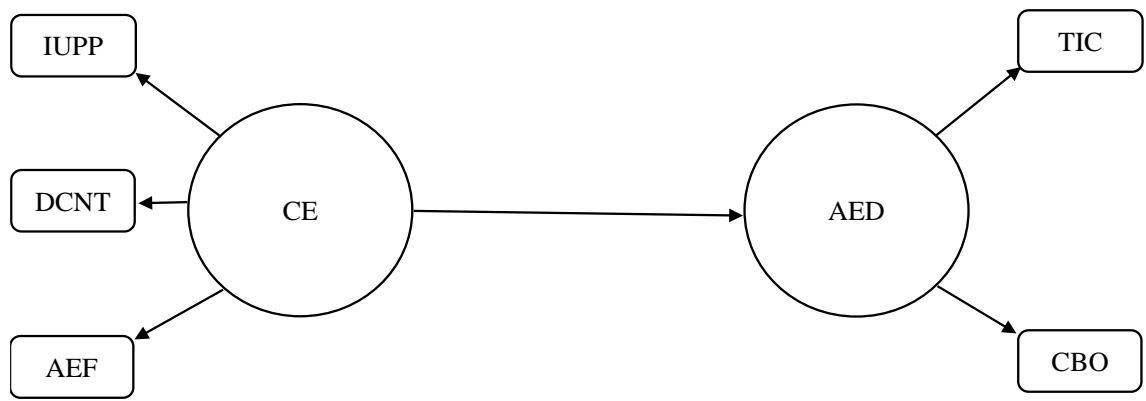

*Model evaluation*

Before starting with the respective multivariate analysis; Hair et al. (2012); Martínez & Fierro (2018a) establish the importance of their evaluation, which implies examining the reliability, internal consistency, convergent, and discriminant validation. These evaluations yielded the following results explained below:

Table 3. Precision and accuracy tests

| Code | Cronbach's α | AVE | PFC | $Q^2$ | HTMT | VIF |
|---|---|---|---|---|---|---|
| CE | .967 | .521 | .969 | | | |
| AED | .988 | .814 | .989 | .494 | | |
| CE-AED | | | | | .844 | 1.00 |

The results obtained show construct reliability in the model since the tests obtained values higher than p-value = .7. Regarding the convergent validation through the test (AVE), we conclude that the set of indicators represents a single underlying construct since values higher than p-value = .50 were obtained (Martínez & Fierro, 2018).

Therefore, each construct explains at least 50% of the variance of the indicators. When evaluating the collinearity level, the test (VIF) did not find problems related to collinearity since its values fluctuated at a p-value = 1.00. In the discriminant validity test or the Heterotrait-Monotrait HTMT methodology, results in less than 0.7 confirm the existence of validity.

The model's predictive quality was performed using the Stone-Geisser redundancy test of cross-validation of the construct or Q2, which assesses the structural and theoretical model; with the results obtained with a value greater than zero 0, the conclusion is drawn existence of predictive validity and relevance of the model (Thaisaiyi, 2020).

*Magnitude and significance of the model*

*Path coefficient results (β) and values (p)*

The analysis of the PLS algorithm's magnitude and significance allows us to measure and test the research model's respective hypothesis relationships. The magnitude is observed in the standardized regression coefficient (β) and its significance (p). With the Bootstrapping algorithm, the magnitude is observed in the standardized regression coefficient (β), and the significance in the two-tailed t (4900) values; where the critical value is (0.01; 4999) = 2,576 (Martínez & Fierro, 2018a). The resampling analysis evaluated (5000 subsamples) with a confidence level of 0.01.

Figure 2. Total effects EC - AED ratio

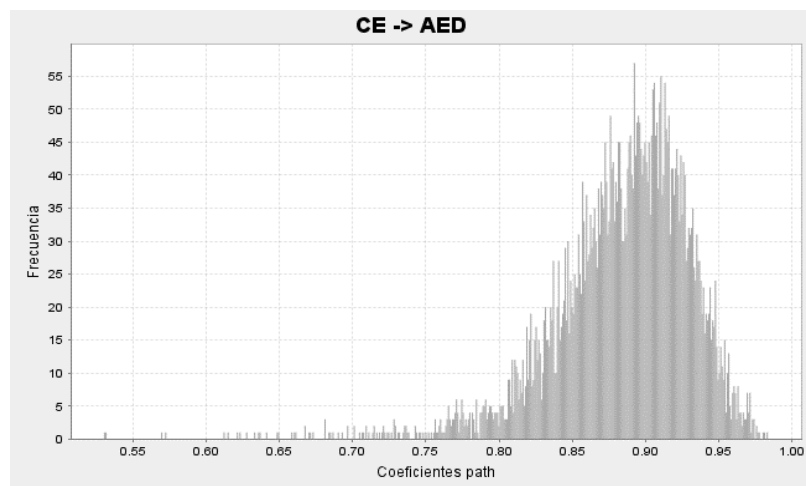

*Results*

The confirmatory analysis of the PLS least squares regression test establishes a high and robust significant impact between CE technological knowledge and AED digital economic activity through a predictive value of $R^2 = .749$, $β = .867$, $p = .000$ (t-value = 18.298). The research showed that for each unit of increase in technological knowledge, digital economic activity increases by 74.9%.

Table 4. Hypothesis test results

| Hypothetical Relationship | Coefficient β | t Student Boostrapping | $R^2$ | p | Hypothesis Accept |
|---|---|---|---|---|---|
| CE-----AED | .867 | 18.298 | .749 | .000 | YES |

*Second analysis phase: microregional spatial regression*

To model the relationship between technological knowledge and digital economic activity, a geographically weighted regression test (GWR) was carried out, which will allow identifying patterns and detecting clusters (Leong & Yue, 2017). It will also allow us to observe its constancy in time and significance (Sabogal, 2013).

This model detects outliers that violate the assumptions that give rise to false correlations (Wilcox, 2016), which prevents the identification of significant values. Consequently, the Montes Claros Microregion was eliminated as it had outlier values. Similarly, the Janauba Microregion was eliminated since no significant relationship was found between the study variables.

Table 5. Regional Comparative Matrix 2009

| Region | IUPP - TIC | IUPP - CBO | DCNT - CBO | EFU - TIC |
|---|---|---|---|---|
| | $R^2$ \| $P$ | $R^2$ \| $P$ | $R^2$ \| $P$ | $R^2$ \| $P$ |
| SALINAS | | .522 \| .002 | | |
| PIRAPORA | .580 \| .017 | .989 \| .000 | .619 \| .014 | |
| BOCAIUVA | .000 \| .000 | .000 \| .000 | .829 \| .023 | |
| G. MOGOL | .000 \| .000 | .000 \| .000 | .000 \| .000 | .921 \| .025 |
| JANAURIA | | | .901 \| .000 | .402 \| .005 |

Table 6. Regional Comparative Matrix 2018

| Region | IUPP - TIC | IUPP - CBO | DCNT - CBO | EFU - TIC |
|---|---|---|---|---|
| | $R^2$ \| $P$ | $R^2$ \| $P$ | $R^2$ \| $P$ | $R^2$ \| $P$ |
| SALINAS | .712 \| .000 | .916 \| .000 | .772 \| .000 | |
| PIRAPORA | | .993 \| .000 | .995 \| .000 | .973 \| .044 |
| BOCAIUVA | .978 \| .001 | .940 \| .006 | .940 \| .014 | .972 \| .042 |
| G. MOGOL | .000 \| .000 | .000 \| .000 | .000 \| .000 | .645 \| .002 |
| JANUARIA | | | .981 \| .000 | |

*Results*

The results confirm the predictive relationship and hypothesis validation between CE technological knowledge and AED digital economic activity in the following microregions:

- 18 Municipalities of the Microregion of Salinas among the IUPP - CBO; IUPP - TIC and EFU - TIC indicators.
- 9 Municipalities of the Pirapora Microregion among the IUPP - CBO; DCNT-CBO; IUPP - DCNT and IUPP - TIC indicators.
- 5 Municipalities of the Bocaiuva Microregion among the IUPP - CBO; IUPP-CBO; DCNT - CBO and EFU - TIC indicators.
- 6 Municipalities of the Microregion of Grao Mogol among the EFU-TIC indicators.
- 16 Municipalities of the Microregion of Juanuaria between the DCNT - CBO and EFU - TIC indicators.

***Third phase of analysis: municipal growth analysis patterns by variable***

The previously geographically weighted regression analysis results suggest the need to distinguish the geographical concentration caused by the phenomena under study (Prat & Cànoves, 2014). Therefore, with the Geoda program, municipalities' growth and concentration patterns were mapped using a statistical grouping technique. The findings identified concentration levels of> 75%.

Table 7. Spatial matrix of similarities at the regional level variable IUPP

| IUPP 2009 (>75%) | IUPP 2018 (>75%) |
|---|---|
| 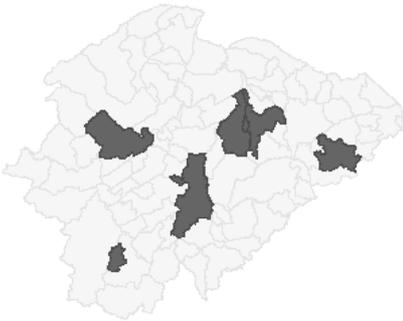 | 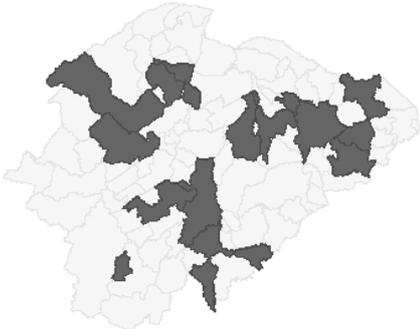 |
| **Municipalities:** San Francisco; Pirapora; Montes Claros; Janauba; Nova Porterinha; Porterinha; Salinas | **Municipalities:** Januária; San Francisco; Pirapora; Montes Claros; Corazón de Jesús; Bocaiúva; Janauba; Mato Verde; Porterinha; Salinas; Rio Pardo de Minas; Taiobeiras; Sao Joao do Paraíso; Itacarambi |

Table 8. Spatial matrix of similarities at the regional level variable DNCT

| DNCT 2009 (>75%) | DNCT 2018 (>75%) |
|---|---|
| 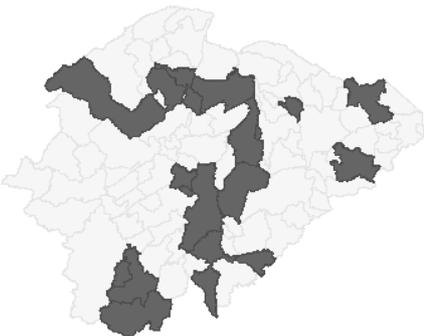 | 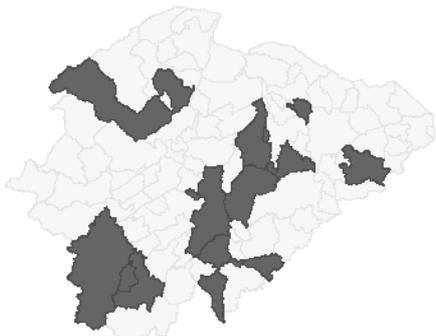 |
| **Municipalities:** Januária; Pirapora; Mirabela; Montes Claros; Bocaiúva; Jaiba; Francisco Sa; Janauba; Nova Porterinha; Mato Verde; Salinas; Sao Joao do Paraíso; Itacarambi; Lassance; Varzea da Palma | **Municipalities:** Januária; Pirapora; Montes Claros; Bocaiúva; Francisco Sa; Janauba; Nova Porterinha; Riacho dos Machados; Mato Verde; Salinas; Buritizeiro; Varzea da Palma |

Table 9. Spatial matrix of similarities at the regional level variable AEF

| AEF 2009 (>75%) | AEF 2018 (>75%) |
|---|---|
| 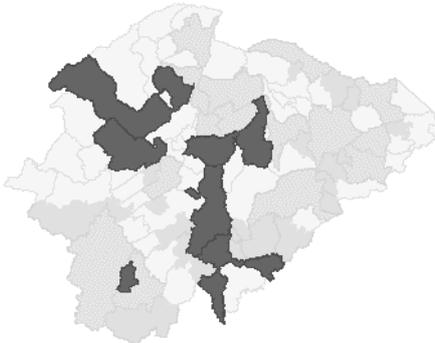 | 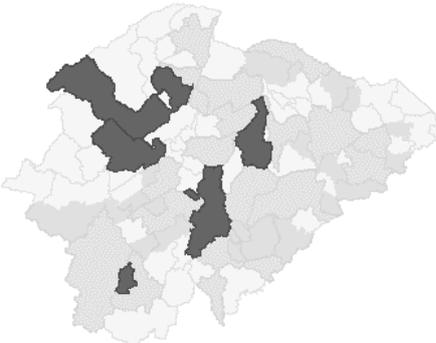 |
| **Municipalities:** Januária; San Francisco; Pirapora; Sao Joao da Ponte; Montes Claros; Bocaiuva; Janauba | **Municipalities:** Januária; San Francisco; Pirapora; Montes Claros; Janauba |

Table 10. Spatial matrix of similarities at the regional level variable TIC

| TIC 2009 (>75%) | TIC 2018 (>75%) |
|---|---|
| 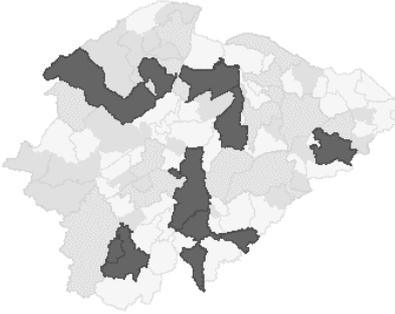 | 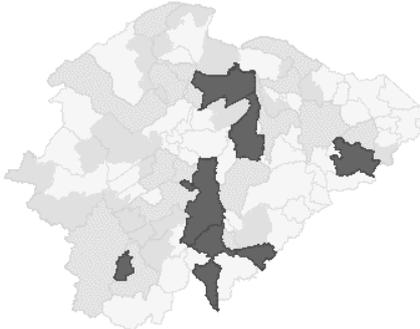 |
| **Municipalities:** Janauria; Pirapora; Montes Claros; Bocaiuva; Jaiba; Janauba; Salinas; Varzea da Palma | **Municipalities:** Janauba; Pirapora; Montes Claros; Bocaiuva; Jaiba; Salinas |

Table 11. Spatial matrix of similarities at the regional level variable CBO

| CBO 2009 (>75%) | CBO 2018 (>75%) |
|---|---|
| 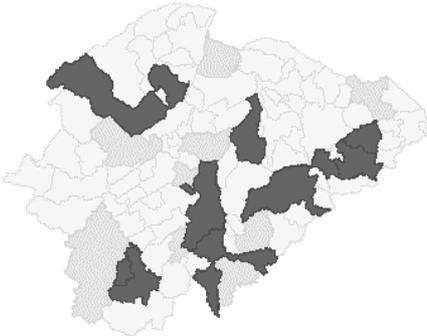 | 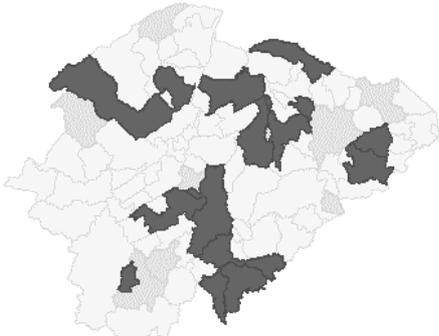 |
| **Municipalities:** Janauria; Pirapora; Montes Claros; Bocaiuva; Janauba; Grao Mogol; Fruta de Leite; Salinas; Taiobeiras; Varzea da Palma | **Municipalities:** Janauria; Pirapora; Corazon de Jesus; Montes Claros; Bocaiuva; Jaiba; Olhos de Agua; Janauba; Espinosa; Porteirinha; Mato Verde; Salinas; Taiobeiras |

The geographic concentration caused by the study phenomena identified several that were statistically analyzed by creating binary dummy variables to be used through the supervised learning model with the use of logistic regression. The Municipalities with the identifier (1) reached values higher than 75% of their neighbors' growth. The municipalities that did not reach 75% growth were categorized with an identifier (0).

The use of machine learning requires the validation of the parameters to identify the appropriate model; therefore, the data was divided into two parts, 85% of data for training, and 15% for testing. To test the model, we used Deep Neural Network, Logistic Regression, and Tree; after carrying out the respective cross-validation, we identified that the logistic regression scores were the highest 0.83 AUC; Therefore it was the model to be used to test the model the remaining 15% of data was used. Finally, the evaluation of the logistic regression model is presented below:

Table 12. Machine Learning evaluation model

| | Accuracy | F value | Precision | Recall | Coefficient Phi | Lift | Value K-S | Kendall Tau | Spearman Rho | FPR |
|---|---|---|---|---|---|---|---|---|---|---|
| Values | 50% | .66 | 66.7% | 66.7% | -.33 | 88.9% | 66.7% | .235 | .258 | 100% |

Note: The regression model's evaluation allowed us to measure and compare its performance to evaluate predictions of new instances that the model has never seen before.

Table 13. Machine Learning model prediction results.

| Predictors | Prediction >75% | Prediction <75% |
|---|---|---|
| San Francisco | 0.94998 | 0.02226 |
| Mato Verde | 0.80181 | 0.17121 |
| Salinas | 0.7727 | 0.20008 |
| Riacho dos Machados | 0.73077 | 0.23663 |
| Nova Porterinha | 0.72089 | 0.24927 |
| Jaiba | 0.70734 | 0.2565 |
| Buritizeiro | 0.6555 | 0.31602 |
| Porterinha | 0.64916 | 0.32188 |
| Bocaiuva | 0.64916 | 0.32188 |
| Rio Pardo do Minas | 0.64916 | 0.32188 |
| Sao Joao do Paraiso | 0.64916 | 0.32188 |
| Varzea da Palma | 0.64916 | 0.32188 |
| Corazon de Jesus | 0.64874 | 0.32167 |
| Itacarambi | 0.64792 | 0.32126 |
| Taiobeiras | 0.63943 | 0.33081 |
| Sao Joao da Ponte | 0.63383 | 0.31427 |
| Januária | 0.63187 | 0.34119 |
| Grao Mogol | 0.45364 | 0.52389 |
| Fruta de Leite | 0.4409 | 0.51521 |

Figure 3. Evaluation curves for classification models

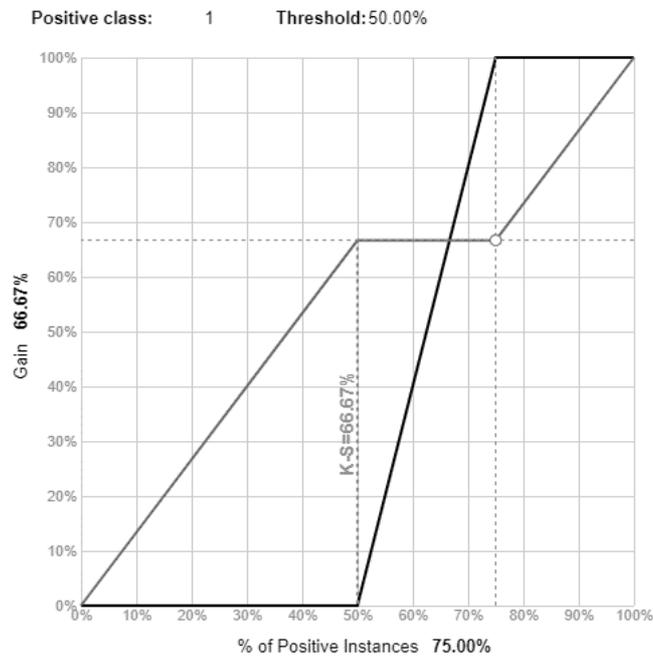

*Results*

The results through machine learning found high predictive values concerning the study variables in 17 municipalities where Sao Francisco (Januária) stood out with 0.94, Mato Verde (Janauba) with 0.80, Salinas (Salinas) with 0.77, Buritizeiro (Pirapora) with 0.65; Bocaiuva (Bocaiuva) with 0.64. However, in Grao Mogol, a predictive value of 0.45 was found, and Fruta de Leite one of 0.44. A side-by-side bar graph is presented to explain better these findings, which compares each municipality with its predictive value.

**Discussion**

The results of this study show that the regions that took advantage of skilled labor based on technological innovation (Manesh, Pellegrini & Marzi, 2019) maintained a more significant competitive advantage over their neighbors and minimized dependence on technological externalities (Agrawal et al., 2019).

These regions have reevaluated their existing strategies and operations to emphasize their objectives of building regional economies based on innovation and technological knowledge (Mendez, 2016), as is the case of the Municipalities of Montes Claros and Pirapora with high results in academic investment and not traditional. Productive processes, quite the opposite with the Municipality of Grao Mogol.

The causes that contributed to the findings in this case study, the gross value added (GVA) of Grao Mogol broke down, and limited investment was discovered in the education and service industry sectors, which generated unfavorable economic results about others. Municipalities such as Montes Claros and Pirapora.

Table 14. Gross value-added matrix (GVA)

| **Municipality** | **2017 Agropec.** | **2009 Agropec.** | **2017 Educ/other** | **2009 Educ/other** | **2017 Industry** | **2009 Industry** |
|---|---|---|---|---|---|---|
| Grao Mogol | 82,836.31 | 8,855 | 43,569.44 | 18,827 | 84,205.77 | 167,114 |
| Pirapora | 30,134.29 | 17,800 | 618,643,59 | 251,077 | 586,648.81 | 311,239 |
| Montes Claros | 112,319.70 | 78,625 | 4,769.729.31 | 2,041,404 | 1,556.646.34 | 879,817 |

Note: Analysis of three cities in the north of Minas Gerais (Unit: R $ x1000)

Figure 4. Gross added value at current prices of economic activity

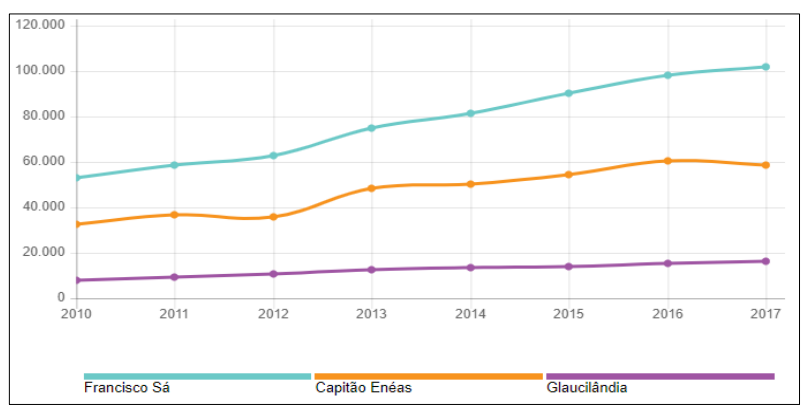

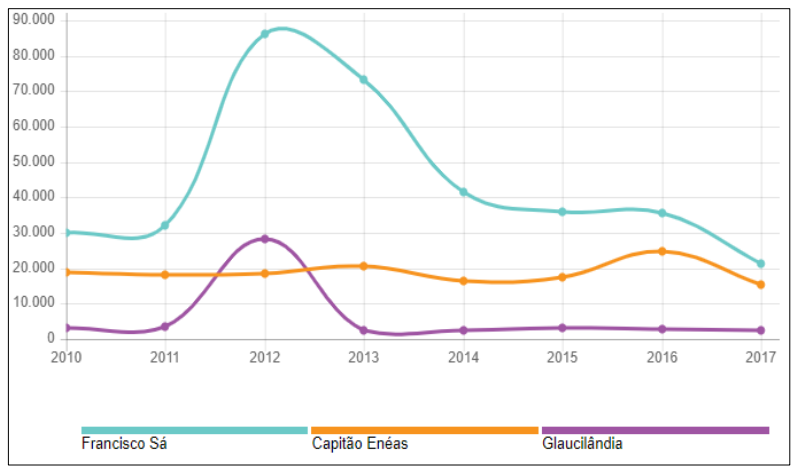

Note: Non-traditional (left); Traditional (right), unit: R $ x1000

In the traditional production model used in the Municipality of Grao Mogol, it was possible to identify two determining factors that respond to this study's need: the low level of schooling of employers and workers and the minimal investment in knowledge technological education. These factors negatively impacted the development of digital economic activity in this region, which is consistent with what was stated by Brynjolfsson et al. (2018); dos Santos (2017); Gomez (2017); Mendez (2016); Meijer & Bolívar (2016); Century et al. (2007) & Cortright (2001).

The evidence also showed that in regions where there is a higher concentration of labor and qualified suppliers in technological innovation, digital economic activity receives a significant boost promoting regional poles of economic development (PRDE) as argued by Sabogal (2013); Siabato & Guzmán-Manrique (2019).

Figure 5. Gross added value at current prices of economic activity

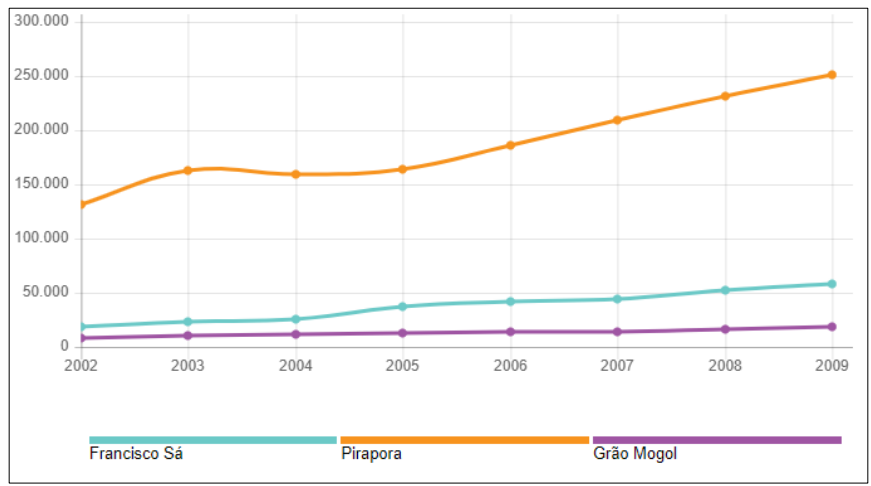

Note: education activity with other services, unit: R $ x1000

Figure 6. Gross added value at current prices of economic activity

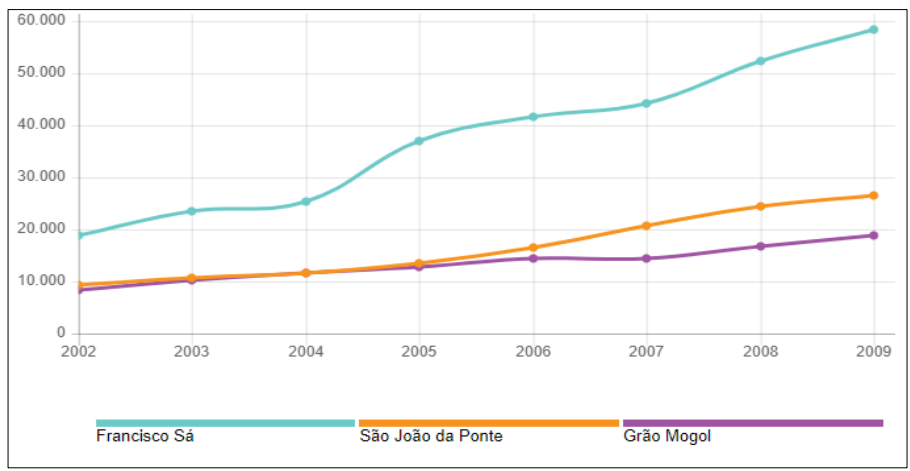

Note: education activity with other services, unit: R $ x1000

The second-order contagion effect was another critical finding in this study, noting that the Municipality of Bocaiúva was favored by its proximity to the Municipalities of Montes Claros and Pirapora. Proving in this way what was argued by Sabogal (2013); Siabato & Guzmán-Manrique (2019) affirm that there may be results in their neighbors in the contagion effect because companies tend to focus on geographic locations in order to use highly qualified labor and suppliers.

Figure 7. Gross added value at current prices of economic activity

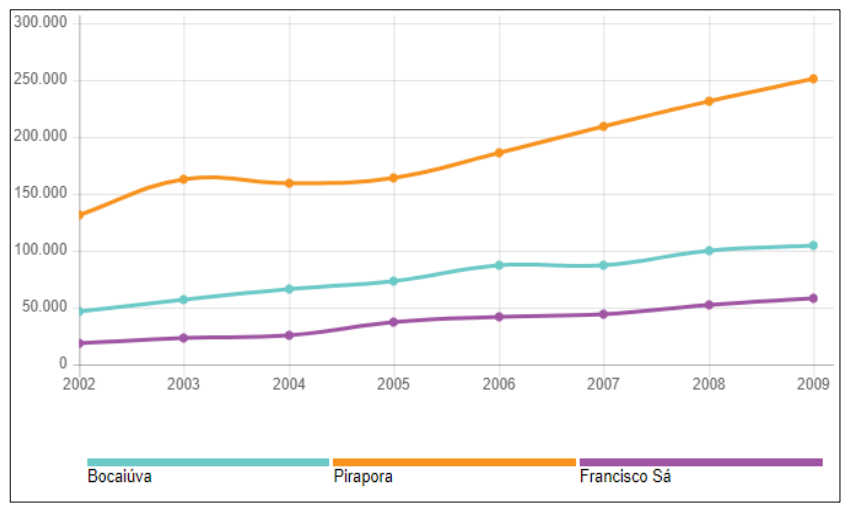

Note: education activity with other services, unit: R $ x1000

**Conclusions**

The theory of economic growth proposed in this study made it possible to identify and test the causal relationship of the variables that establish that the production level increases due to the transfer of knowledge in human capital and information technologies (Agrawal et al., 2019).

This article also provides new evidence that contrasts with arguments that propose urban decentralization as a factor that would alleviate urban pressures on large cities by promoting small cities (Lu et al., 2018) due to the contagion effect of cities with economies based on innovation and technological knowledge affects services, production, and marketing, generating new opportunities in the labor market.

Consequently, municipal governments must initiate a transformation framed in a new governmental, technological management (a subject proposed in this study for future research). This management model would help develop new business models linked to technological innovation through specialized technological training, preventing the effects that slow down the development of small municipalities due to the geographic concentration of companies that demand technology-based skills.